\documentstyle[12pt]{article}

\input epsf
 \begin{document}

\begin{center}
\large{{\bf A Machian Model of Dark Energy}}
\end{center}

\begin{center}
R. G. Vishwakarma

{\it Inter-University Centre for Astronomy \& Astrophysics,
 Post Bag 4, Ganeshkhind, Pune 411 007, India }\\
 E-mail: vishwa@iucaa.ernet.in
\end{center}

\vspace{1.5cm}

\begin{abstract}
\noindent
Einstein believed that Mach's principle should play a major role in
finding a meaningful spacetime geometry, though it was discovered later
that his field equations gave some solutions which were not Machian. It is
shown, in this essay, that the kinematical $\Lambda$ models, which are
invoked to solve the cosmological constant problem, are in fact
consistent with Mach's ideas. One particular model in this category is
described which results from the microstructure of spacetime and seems
to explain the current observations successfully and also has some benefits
over the conventional models. This forces one to think whether the Mach's
ideas and the cosmological constant are interrelated in some way.


\vspace{.3cm}

\end{abstract}

\vspace{2cm}
\noindent
Received an {\it Honorable mention} in the Essay Contest$-$2002
sponsored by the {\sf Gravity Research Foundation}.

\newpage

\vspace{1cm}
\noindent
When Einstein introduced the cosmological constant $\Lambda$ into his
field equations to obtain a static solution, he was guided by Mach's
principle, which argued that the distribution of matter determined the
precise geometrical nature of spacetime and hence forbade the notion
of empty universe. He believed that the presence of matter was
essential for a meaningful spacetime geometry [1]. However, he had to
discount his idea when deSitter discovered a cosmological model
with $\Lambda$ and no matter at all which had both static and dynamic
representations. Later he also dismissed $\Lambda$ when it was found that
the universe was expanding.
One however notices that if a dynamic $\Lambda (t)$ is introduced
into Einsein's field equations, no solution is possible in the absence
of matter. This is clear from the divergence of the field equations:

\begin{equation}
[R^{ij}-\frac{1}{2} R g^{ij}]_{;j}=0=-8\pi G\left[T^{ij}-
\frac{\Lambda (t)}{8\pi G} ~ g^{ij}\right]_{;j}.
\end{equation}
(I shall use the units with $c=1$ throughout. However, $c$ will be restored
whenever needed.)
Obviously a solution with a dynamic $\Lambda$ is possible only if
$T^{ij} \neq 0$ (and $T^{ij}_{;j} \neq 0$). In the absence of matter
(or even if the matter is conserved), $\Lambda$ has got to remain a
constant. Thus the empty spacetime cannot be obtained as a solution of
general relativity with a {\it dynamic $\Lambda (t)$}.

This way of introducing $\Lambda$ into Einstein's equations gives
it a status of a source term. Now it ($\Lambda/8\pi G$) represents the
energy density of `{\it emptiness}' (vacuum) and hence invites particle
physics to interact with general relativity via $\Lambda$. Note that the only possible
covariant form for the energy momentum tensor of the quantum vacuum is
$T^{ij}_{\rm v} = -\rho_{\rm v} g^{ij}$, which is equivalent to the
cosmological constant. It behaves like a perfect fluid with the energy
density $\rho_{\rm v}=\Lambda/8\pi G$ and an isotropic pressure
$p_{\rm v}=-\rho_{\rm v}=-\Lambda/8\pi G$.
The conserved quantity is now the sum of matter and vacuum (and not the
two separately), as is obvious from equation (1).

Here comes the problem: the value of vacuum energy at the Planck epoch comes 
out as $\approx 10^{76}$ GeV$^4$, which is 123 orders of magnitude
larger than its value predicted by the Friedmann equation

\begin{equation}
\frac{\dot S^2}{S^2}+\frac{k}{S^2}=\frac{8\pi G}{3}\rho+\frac{\Lambda}{3},
\end{equation}
which gives $\Lambda_0\approx H_0^2$ or equivalently $\rho_{\rm v0}\approx 10^{-47}$
GeV$^4$ [2]. (The subscript `0' denotes the value of the quantity at the
present epoch.)
It is fortunate for general relativity that this predicted value of $\Lambda$
by the theory  is also consistent with the recent observations of type
Ia supernovae [3] and the anisotropy measurements of the cosmic microwave
background radiation (CMBR) [4], taken together with the complimentary observational
constraints on matter density [5]; all indicate that the present constituent
of the universe is dominated by some weird kind of energy with negative
pressure, commonly known as `{\it dark energy}'. The simplest candidate
for {\it dark energy} is the cosmological constant, though plagued with
this so called the {\it cosmological constant problem}.
Obviously the problem arises due to the incompatibility of general
relativity and particle physics.
The dynamical $\Lambda$ was, in fact, invoked in an attempt
(phenomenological in nature) to solve this problem
(historically, it was not invoked to make the solutions of Einsein's
equations consistent with Mach's ideas). The rationale behind this approach
is that $\Lambda$ was large during the early epochs and it decayed as the
universe evolved, reducing to a small value at the present epoch.

There is another phenomenological approach to solve this problem,
which has become very
popular since recent observations suggested the existence of a nonzero 
$\Lambda$. This invokes a slowly rolling down scalar field $\phi$, commonly
known as `{\it quintessence}', with an appropriate potential $V(\phi)$ to
explain the observations [6].

Note that though the quintessence fields also acquire negative
pressure during the matter dominated phase and behave like  dynamical $\Lambda$
(with $\Lambda_{\rm effective}\equiv 8\pi G\rho_{\phi}$), they are in
general fundamentally different from the dynamical (kinematical) $\Lambda$.
In the former case, quintessence and matter
fields are assumed to be conserved separately (through the assumption of
minimal coupling of the scalar field with the matter fields). However, in the
latter case, the conserved quantity is $[T^{ij} + T^{ij}_{\rm v}]$,
as have been mentioned earlier. This implies that there is a continuous creation
of matter from the {\it decaying} $\Lambda$ as is clear from the following.

\begin{equation}
\rho=C S^{-3(1+\omega)}-\frac{S^{-3(1+\omega)}}{8\pi G}
\int \dot\Lambda(t) S^{3(1+\omega)}{\rm d}t, ~ ~ ~ C=\mbox{constant},
\end{equation}
which follows from (1) and suggests that there is a positive contribution to
$\rho$ from the {\it decaying} $\Lambda$ ($\dot \Lambda<0$). Here
$\omega=p/\rho$ is the usual equation of state of the matter field.
Obviously the quintessence models need not be consistent with Mach's ideas.

It may also be noted that, for a given pair of $S(t)$ and $\rho(t)$, it is
always possible to find a $V(\phi)$ which explains the observations,
as has been shown recently by Padmanabhan [7]. This result
is irrespective of what the future observations reveal about the given $S(t)$
and $\rho(t)$, and hence makes these models trivial. Like the anthropic
principle [8], these models also don't have any predictive power and lead to
similar late time behaviour of the universe.

The true solution of the cosmological constant problem should be provided by a
full theory of quantum cosmology, which is unfortunately not available at the
moment. However, some arguments have been made, based on the quantum
gravitational uncertainty principle and the discrete structure of spacetime
at Planck length, which have made it possible to connect the cosmological
constant with the microstructure of spacetime [7, 9]. By assuming that
$\Lambda$ is a stochastic
variable arising from the quantum fluctuations and it is the rms fluctuation
which is being observed in the cosmological context, it has been shown that
the uncertainty in the value of $\Lambda$ can be written as

\begin{equation}
\Delta\Lambda=\frac{1}{\sqrt{\mho}},
\end{equation}
where $\mho$ is the four volume of the universe. If one estimates the
`{\it radius}' of the universe by $S\approx c t \approx c H^{-1}$, then this
reduces to
\begin{equation}
\Delta\Lambda \approx H^2, ~ ~ (\mbox{in units with}~ c=1),
\end{equation}
which matches exactly with the present observations.

There are also other ways which suggest $\Lambda \propto H^2$. Two such
ways have been described in the following (two more have been described
by Padmanabhan in his paper [7]).

(i) We know that a positive $\Lambda$ introduces a force of repulsion between two
bodies which increases in proportion to the distance between them. This force
experienced by a test particle at the scale of the whole universe is
$c \Lambda H^{-1}$. If this repulsive force roughly balances the gravitational
attraction $4\pi G c\rho/3H$ of the universe on the test particle, one finds
$\Lambda \approx H^2$, provided $\Omega_{\rm m}$ ($\equiv 8\pi G \rho/3H^2$) is of order
unity.

(ii)  From the dimensional considerations, it is always possible to write
$\Lambda$ in terms of Planck energy density times a dimensionless quantity
[10]:

\begin{equation}
\Lambda \approx 8\pi G \rho_{\rm Pl}\left[\frac{t_{\rm Pl}}{t_H}
\right]^\alpha \approx t_{\rm Pl}^{-2} \left[\frac{t_{\rm Pl}}{t_H}
\right]^\alpha,
\end{equation}
where $t_{\rm Pl}\equiv (G \hbar/c^5)^{1/2}$ and $t_H \equiv H^{-1}$ are
the Planck and Hubble times respectively and $\rho_{\rm Pl}\equiv c^5/G^2 \hbar$
is the Planck energy density. For $\alpha=2$, which gives the right value of $\Lambda$
at the present epoch, equation (6) leads to $\Lambda \approx H^2$.

By writing this law as $\Lambda =n H^2$, where $n$ is a constant parameter,
the dynamics of the resulting model can be obtained, from equations (2) and
(3), as

\begin{equation}
\rho \propto \Lambda \propto H^2 \propto t^{-2}, ~ S \propto t^{2/[(3-n)(1+\omega)]},
~ n < 3,
\end{equation}
where we have considered $k=0$, as has been suggested by
the recent CMBR observations [4]. The cases $n\geq3$ (where $\rho\leq0$) are
either unphysical or not compatible with $\Lambda=\Lambda(t)$.
Note that the ansatz $\Lambda=n H^2$ is equivalent to assuming that
$\Omega_{\Lambda}$ $(\equiv \Lambda/3 H^2=n/3)$ is a constant and,
hence, so is $\Omega_{\rm m}$
$(=1-n/3)$ in a flat model. Hence $\rho_{\rm v}/(\rho + \rho_{\rm v})=n/3$
is also a constant.
The deceleration parameter, in the model, is obtained as

\begin{equation}
q=\frac{(3-n)(1+\omega)}{2}-1,
\end{equation}
which is also constant and implies that $q\frac{>}{<}0$ according as
$n\frac{<}{>}(1+3\omega)/(1+\omega)$. Thus two different values of the parameter
$n$, viz., one with $n<(1+3\omega)/(1+\omega)$ (say, $n_1$) and the other
with $n>(1+3\omega)/(1+\omega)$ (say, $n_2$) can make the universe
shift from deceleration
to acceleration. This is interesting in view of the result
obtained by Turner and Riess [11],
which shows that the supernovae data favour a past deceleration followed
by a recent acceleration, independent of the content of the universe.
In fact, this is exactly the case in this model, as can be checked from the
constraints on $n$ coming from various observations.
Freese et al [12],
who derived this model by assuming $\rho_{\rm v}/(\rho +
\rho_{\rm v})=$ constant, found that the element abundances from the
primordial nucleosynthesis require $\rho_{\rm v}/(\rho +
\rho_{\rm v})\leq 0.1$. In terms of the parameter $n$, this translates to
an $n\leq 0.3$
in the early radiation era, implying a deceleration.
Let us now see how the present observations constrain the
model. It has already been shown that the model fits the high
redshift supernovae Ia data (including SN 1997ff at $z\approx 1.7$)
very well [13].
Additionally, it also fits the data on the angular size and redshift of the
compact radio sources very well [14]. Both the observations require
$n\approx 1.5$ and hence predict an accelerating expansion at the present
epoch.

Interestingly these constraints are also consistent with the CMBR
anisotropy observations which, especially the first peak in the angular power
spectrum curve which has been confirmed by various observations, require $n$
to change at a redshift of a few.
Thus if the expansion dynamics switches over from deceleration to acceleration
at $z=z_1$, the {\it angular diameter distance} to the last scattering
surface (at $z=z_{\rm dec}$) is given by

\begin{equation}
d_{\rm A} = \frac{1}{(1+z_{\rm dec})}\left[\int_0^{z_1}\frac{dz}{H(n_2;z)}+
\int_{z_1}^{z_{\rm dec}}\frac{dz}{H(n_1;z)}\right].
\end{equation}
For the dynamics of the model given by equation (7), this yields

\begin{equation}
d_{\rm A} = \frac{1}{H_0(1+z_{\rm dec})}\left[\int_0^{z_1} (1+z)^{(n_2-3)/2} dz+
\int_{z_1}^{z_{\rm dec}}(1+z)^{(n_1-3)/2} dz\right].
\end{equation}
If one considers $n_2=1.5$ (from the SN and the radio sources data)
and $z_1=5$ (to be on the safe side in view of the
future higher redshift observations), then a value of $n_1=0.15$ gives
the angle subtended by the Hubble radius $d_{\rm H}(z_{\rm dec},n_1)$
(with $z_{\rm dec}=1100$) at the observer as $\approx 0.9^0$ which is equivalent
to a peak at a Legendre multipole size $\ell \approx 200$. This is exactly what the
CMBR anisotropy observations have measured.
Note that the parameter space ($n_1$, $z_1$) is wide enough which makes the
model robust.

\begin{figure}
\centerline{{\epsfxsize=14cm {\epsfbox[50 250 550 550]{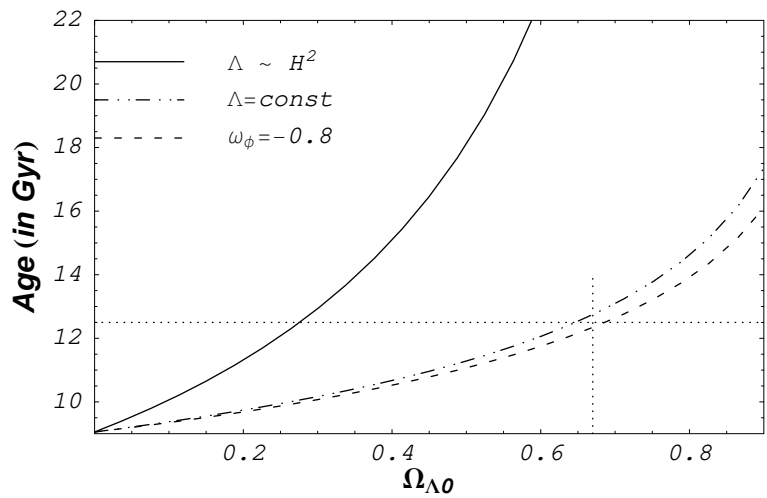}}}}
\caption{ The age of the universe is plotted as a function of
$\Omega_{\Lambda0}$ in some flat models, by using $H_0=72$ km
s$^{-1}$ Mpc$^{-1}$. The horizontal dotted line represents the
age of the globular clusters $t_{\rm GC}=12.5$ Gyr. The
vertical dotted line corresponds to the mass density of the present
universe ($\Omega_{\rm m0}=0.33$).}
\end{figure}    

Another attractive feature of the model is that it supplies a sufficiently
large age of the universe, which is very remarkable in view of the fact that
the age of the universe in the FRW model with a constant $\Lambda$ is
uncomfortably close to the age of the globular clusters $t_{\rm GC}=12.5
\pm 1.2$ Gyr [15].  The {\it quintessential} models give even lower age.
In Figure 1, we have plotted the expansion age of the universe $t_0$ as a
function of $\Omega_{\Lambda0}$ in the present model, together with the
favoured quintessence model ($\omega_{\phi}\equiv p_{\phi}/\rho_{\phi}=-0.8$)
and the FRW model with a constant $\Lambda$ ($\omega_{\phi}=-1$).
Note that if the required mass density $\Omega_{\rm m0}$ of the universe
was smaller, one could get higher age in these models, as is clear from
the figure. This does not, however, seem likely, as the recent measurements
give very narrow range of $\Omega_{\rm m0}$ as $\Omega_{\rm m0}=0.330\pm 0.035$ at
one sigma level [5]. By using $H_0=72\pm 7$ km s$^{-1}$ Mpc$^{-1}$ (which
is recently
measured by the {\it Hubble Space Telescope} key project and is also
consistent with a host of other experiments [16]), this value of
$\Omega_{\rm m0}$ gives $t_0=12.7 \pm 1.6$ Gyr in the FRW model with a constant
$\Lambda$. This is roughly consistent with the value $t_0 = 14 \pm 0.5$ Gyr
estimated from the CMBR observations, which has been claimed to give more
accurate age of the universe [17].
The value of $t_0$ in the favoured quintessence model is obtained as
$t_0= 12.3\pm 1.5$ Gyr, which seems in real trouble in view of
$t_{\rm GC}=12.5 \pm 1.2$ Gyr.
In this connection it is very encouraging that the model $\Lambda\propto H^2$,
where the expression for the age of the universe yields
$t_0 \approx 2/(3\Omega_{\rm m}) H_0^{-1}$, gives $t_0 \approx 27.4\pm 5.6$ Gyr
which is remarkably high.

In light of the successes and achievements stated above of this model,
one is inclined to ask if it is just a matter of coincidence that the model
is consistent with Mach's ideas and at the same time it solves the
cosmological constant problem (at least phenomenologically). Should Mach's
principle play some fundamental role in solving the cosmological constant
problem? Two concepts (Mach's principle and the cosmological constant), invoked
by Einstein, and later dismissed by himself, seem to be unavoidable. Are they
really interlinked in some intricate way? Only the future will answer
these questions.

\medskip
\noindent
{\bf ACKNOWLEDGEMENTS}

\noindent
The author thanks Professor J. V. Narlikar
for useful discussion and the Department of Atomic Energy, India for support
of the {\it Homi Bhabha postdoctoral fellowship}.

\vspace{1.5cm}
\noindent
{\bf REFERENCES:}

\medskip



 \noindent
[1] Narlikar J. V. (2002), {\it An Introduction to Cosmology}, Cambridge
Univ. Press, Camgridge, p 104.

\medskip \noindent
[2] Weinberg, S. (1989) Rev. Mod. Phys. {\bf 61} 1;
Sahni V. and Starobinsky A. (2000) Int. J. Mod. Phys. D {\bf 9} 373.

\medskip \noindent
[3] Perlmutter S. et al (1999), Astrophys. J. {\bf 517}, 565;
 Riess A. G. et al (1998), Astron. J., {\bf 116}, 1009;
Riess A. G. et al (2001) Astrophys. J. {\bf 560}, 49.

\medskip \noindent
[4] de Bernardis P. et al (2000) Nature {\bf 404}, 955;
Lee A. T. et al (2001) Astrophys. J. {\bf 561}, L1; Halverson N. W. et al
(2002) Astrophys. J. {\bf 568}, 38; Sievers J. L. et al, astro-ph/0205387.

\medskip\noindent
[5] Turner M. S., astro-ph/0106035.

\medskip\noindent
[6] Zlatev L., Wang L. and Steinhardt P. J. (1999) Phys. Rev. Lett.
{\bf 82}, 896;
Caldwell P. R., Dave R. and Steinhardt P. J. (1998) Phys. Rev. Lett.
{\bf 80}, 1582;
Wetterich C. (1988) Nucl. Phys. B. {\bf 302}, 668;
Ratra B. and Peebles P. J. E. (1988) Phys. Rev. D. {\bf 37}, 3406;
Peebles P. J. E. and Ratra B. (1988) Astrophys. J. {\bf 325}, L17.

\medskip\noindent
[7] Padmanabhan T., gr-qc/0112068.

\medskip\noindent
[8] Weinberg S., astro-ph/0005265.

 \medskip\noindent
[9] Sorkin R. D. (1997), Int. J. Theor. Phys. {\bf 36}, 2759.

\medskip\noindent
[10] Chen, W. and Wu, Y. S. (1990) Phys. Rev. D {\bf 41}, 695;
 Carvalho, J. C., Lima, J. A. S. and Waga, I. (1992) Phys. Rev. D
 {\bf 46}, 2404.

\medskip\noindent
[11] Turner M. S. and Riess A. G., astro-ph/0106051.

 \medskip\noindent
[12] Freese, K., Adams, F. C., Friemann, J. A. and Mottolla E. (1987)
  Nucl.Phys. B, {\bf 287}, 797.

\medskip\noindent
[13] Vishwakarma, R. G. (2001)  Class. Quantum Grav. {\bf 18},1159;
 Vishwakarma, R. G. (2002) MNRAS, {\bf 331}, 776. 

\medskip\noindent
[14] Vishwakarma, R. G. (2000)  Class. Quantum Grav. {\bf 17}, 3833.

\medskip\noindent
[15] Gnedin O. Y., Lahav O., Rees M. J., astro-ph/0108034;
Cayrel R., et al, (2001), Nature, {\bf 409}, 691.

\medskip\noindent
[16] Freedman W. L. (2001), Astrophys. J. {\bf 553}, 47;
Turner M. S., astro-ph/0202008.

\medskip\noindent
[17] Knox L. and Skordis C., astro-ph/0109232.

\end{document}